\documentstyle[aps,graphicx,epsf]{revtex}\tighten
\draft

\def \D {\mbox{D}}

\begin{document}
\twocolumn[\hsize\textwidth\columnwidth\hsize\csname
@twocolumnfalse\endcsname

\title{Anisotropy dissipation in brane-world inflation}

\author{Roy Maartens$^{1,*}$, Varun Sahni$^{2,\dagger}$ and Tarun Deep
Saini$^{2,\ddagger}$} \address{$^1$Relativity and Cosmology Group,
School of Computer Science and Mathematics, Portsmouth University,
Portsmouth~PO1~2EG, Britain} \address{$^2$Inter-University Centre for
Astronomy and Astrophysics, Ganeshkhind, Pune 411 007, India}

\maketitle

\begin{abstract}

We examine the behavior of an anisotropic brane-world in the
presence of inflationary scalar fields. We show that, contrary to
naive expectations, a large anisotropy does not adversely affect
inflation. On the contrary, a large initial anisotropy introduces
more damping into the scalar field equation of motion, resulting
in greater inflation. The rapid decay of anisotropy in the
brane-world significantly increases the class of initial
conditions from which the observed universe could have originated.
This generalizes a similar result in general relativity.
A unique feature of Bianchi I brane-world cosmology appears to be that
for scalar fields with a large kinetic term the initial expansion
of the universe is {\em quasi-isotropic}.
The universe grows more anisotropic during an intermediate
transient regime until anisotropy
finally disappears during inflationary expansion.
\end{abstract}

 \vskip2pc]

\section{Introduction}

Observations of galaxies, QSO's and the cosmic microwave
background appear to indicate that we live in a universe which is
remarkably uniform on very large scales. Yet the homogeneity and
isotropy of the universe is difficult to explain within the
standard relativistic framework since, in the presence of matter,
the class of solutions to the Einstein equations which evolve
towards a FRW universe is essentially a set of measure
zero~\cite{ch73}. The above statement is however only true for
space-times containing `normal' matter satisfying `energy
conditions' which ensure that (i) negative pressures can never
grow so large as to dominate the energy density: $T_{00} \geq
\vert T_{ij}\vert$; (ii) the sum of the principle pressures of the
fluid must be non-negative: $\sum_{i=1}^3 T_{ii} \geq 0$. The
inflationary scenario, based as it is on a form of matter which
violates these energy conditions, radically alters the above
picture. Indeed, as demonstrated in~\cite{wald,star83,ms86,tw86,js86,rellis86}, a
large class of spacetimes both homogenize and isotropize under the
influence of an effective cosmological $\Lambda$-term. Thus the
inflationary scenario can successfully generate a homogeneous and
isotropic FRW-like universe from initial conditions which, in the
absence of $\Lambda$, would have resulted in a universe far
removed from the one we live in today.

Recently there has been a great deal of interest in a cosmological
scenario in which matter fields are confined to a 3 dimensional
`brane-world' embedded in a higher dimensional `bulk'
space~\cite{rs}. This higher-dimensional cosmology generalizes the
standard Kaluza-Klein picture by allowing the presence of large or
even infinite non-compact extra dimensions. The issue of inflation
on the brane was investigated in~\cite{mwbh}, where it was shown
that on an FRW brane in 5-dimensional anti de Sitter space,
extra-dimensional effects are conducive to the advent of inflation
(see also~\cite{cll}). In this paper we address the kindred issue
of anisotropic initial conditions. We demonstrate that even very
large initial anisotropy cannot prevent brane-world inflation from
occurring, thus generalizing a previous result in general
relativity~\cite{ms86,tw86,js86,rellis86}.
On the contrary, for a large class of
initial conditions, the presence of anisotropy actually {\em
increases} the amount of inflation. Thus a scalar field dominated
universe can eventually isotropise and inflate, even if its
expansion was very anisotropic to begin with.
A unique feature of brane cosmology is that the effective equation of state
at high densities can become ultra stiff. Consequently
matter can overwhelm shear
for equations of state which are
stiffer than dust, leading to quasi-isotropic early expansion of
the universe in such cases.

\section{Brane dynamics}

The 5-dimensional (bulk) field equations are
\begin{equation}
\widetilde{G}_{AB} =
\widetilde{\kappa}^2\left[-\widetilde{\Lambda}\widetilde{g}_{AB}
+\delta(y)\left\{ -\lambda g_{AB}+T_{AB}\right\}\right]\,,
\label{1}
\end{equation}
where tildes denote the bulk generalization of standard general
relativity quantities, and $\widetilde{\kappa}^2=
8\pi/\widetilde{M}_{\rm p}^3$, where $\widetilde{M}_{\rm p}$ is
the fundamental 5-dimensional Planck mass, which is typically much
less than the effective Planck mass on the brane, $M_{\rm
p}=1.2\times 10^{19}$ GeV. The brane is given in Gaussian normal
coordinates $x^A=(x^\mu,y)$ by $y=0$, where $x^\mu=(t,x^i)$ are
spacetime coordinates on the brane. The brane tension is
$\lambda$, and $g_{AB}=\widetilde{g}_{AB}-n_An_B$ is the induced
metric on the brane, with $n_A$ the space-like unit normal to the
brane. Standard-model matter fields confined to the brane make up
the brane energy-momentum tensor $T_{AB}$ (with $T_{AB}n^B=0$).
The bulk cosmological constant $\widetilde{\Lambda}$ is negative,
and is the only 5-dimensional source of the gravitational field.

The field equations induced on the brane are derived via an
elegant geometric approach in~\cite{sms}, leading to new terms
that carry bulk effects onto the brane:
\begin{equation}
G_{\mu\nu}=-\Lambda g_{\mu\nu}+\kappa^2
T_{\mu\nu}+\widetilde{\kappa}^4S_{\mu\nu} - {\cal E}_{\mu\nu}\,.
\label{2}
\end{equation}
Here $\kappa^2=8\pi/M_{\rm p}^2$ and
\begin{equation}
\lambda=6{\kappa^2\over\widetilde\kappa^4} \,, ~~ \Lambda =
{\textstyle{1\over2}}\widetilde\kappa^2\left(\widetilde{\Lambda}+
{\textstyle{1\over6}}\widetilde\kappa^2\lambda^2\right)\,.
\label{3}
\end{equation}
We assume that $\widetilde{\Lambda}$ is chosen so that
$\Lambda=0$. Extra-dimensional corrections to the Einstein
equations on the brane are of two forms: firstly, the matter
fields contribute local quadratic energy-momentum corrections via
the tensor $S_{\mu\nu}$, and secondly, there are nonlocal effects
from the free gravitational field in the bulk, transmitted via the
projection ${\cal E}_{\mu\nu}$ of the bulk Weyl tensor. The matter
corrections are given by
\begin{eqnarray*}
S_{\mu\nu}&=&{\textstyle{1\over12}}T_\alpha{}^\alpha T_{\mu\nu}
-{\textstyle{1\over4}}T_{\mu\alpha}T^\alpha{}_\nu\nonumber\\
&&~~{}+ {\textstyle{1\over24}}g_{\mu\nu} \left[3 T_{\alpha\beta}
T^{\alpha\beta}-\left(T_\alpha{}^\alpha\right)^2 \right]\,,
\end{eqnarray*}
and are significant at high energies, i.e., $\rho \gtrsim
\lambda$.  The projection of the bulk Weyl tensor is
\[
{\cal E}_{AB}=\widetilde{C}_{ACBD}n^C n^D\,,
\]
which is symmetric and traceless and without components orthogonal
to the brane, so that ${\cal E}_{AB}n^B=0$ and ${\cal E}_{AB}\to
{\cal E}_{\mu\nu}g_A{}^\mu g_B{}^\nu$ as $y\to 0$.

The Weyl tensor $\widetilde{C}_{ABCD}$ represents the free,
nonlocal gravitational field in the bulk, i.e., the part of the
field that is not directly determined at each point by the
energy-momentum tensor at that point. The local part of the bulk
gravitational field is the Einstein tensor $\widetilde{G}_{AB}$,
which is determined locally via the bulk field equations
(\ref{1}). Thus ${\cal E}_{\mu\nu}$ transmits nonlocal
gravitational degrees of freedom from the bulk to the brane,
including tidal (or Coulomb), gravito-magnetic and transverse
traceless (gravitational wave) effects.

If $u^\mu$ is the 4-velocity comoving with matter (which we assume
is a perfect fluid or minimally-coupled scalar field), the
nonlocal term has the form of a radiative energy-momentum
tensor~\cite{m}:
\[
{\cal E}_{\mu\nu}={-6\over\kappa^2\lambda}\left[{\cal
U}\left(u_\mu u_\nu+{\textstyle {1\over3}} h_{\mu\nu}\right)+{\cal
P}_{\mu\nu}+{\cal Q}_{\mu}u_{\nu}+{\cal Q}_{\nu}u_{\mu}\right]\,,
\]
where $h_{\mu\nu}=g_{\mu\nu}+u_\mu u_\nu$ projects into the
comoving rest-space. Here
\[
{\cal U}=-{\textstyle{1\over6}}\kappa^2 \lambda\, {\cal
E}_{\mu\nu}u^\mu u^\nu
\]
is an effective nonlocal energy density on the brane (which need
not be positive), arising from the free gravitational field in the
bulk. It carries Coulomb-type effects from the bulk onto the
brane. There is an effective nonlocal anisotropic stress
\[
{\cal P}_{\mu\nu}=-{\textstyle{1\over6}}\kappa^2 \lambda\left[
h_\mu{}^\alpha h_\nu{}^\beta-{\textstyle{1\over3}}h^{\alpha\beta}
h_{\mu\nu}\right] {\cal E}_{\alpha\beta}
\]
on the brane, which carries Coulomb, gravito-magnetic and
gravitational wave effects of the free gravitational field in the
bulk. The effective nonlocal energy flux on the brane,
\[
{\cal Q}_\mu ={\textstyle{1\over6}}\kappa^2 \lambda\,
h_\mu{}^\alpha {\cal E}_{\alpha\beta}u^\beta\,,
\]
carries Coulomb and gravito-magnetic effects from the free
gravitational field in the bulk.

The local and nonlocal bulk modifications may be consolidated into
an effective total energy-momentum tensor:
\begin{equation}
G_{\mu\nu}=-\Lambda g_{\mu\nu}+\kappa^2 T^{\rm tot}_{\mu\nu}\,,
\label{6'}
\end{equation}
where
\[
T^{\rm tot}_{\mu\nu}= T_{\mu\nu}+{6\over \lambda}S_{\mu\nu}-
{1\over\kappa^2}{\cal E}_{\mu\nu}\,.
\]
The effective total energy density, pressure, anisotropic stress
and energy flux are~\cite{m}
\begin{eqnarray}
\rho^{\rm tot} &=& \rho\left(1+{\rho\over2\lambda}\right)+{6 {\cal
U}\over\kappa^4\lambda}\,, \label{a}\\ p^{\rm tot} &=& p+
{\rho\over2\lambda}(\rho+2p) +{2{\cal U}\over\kappa^4\lambda}\,,
\label{b}\\ \pi^{\rm tot}_{\mu\nu} &=&{6\over
\kappa^4\lambda}{\cal P}_{\mu\nu}\,, \label{c}\\ q^{\rm tot}_\mu
&=& {6\over \kappa^4\lambda}{\cal Q}_\mu \,.\label{d}
\end{eqnarray}

The brane energy-momentum tensor separately satisfies the
conservation equations, $\nabla^\nu T_{\mu\nu}=0 $. The Bianchi
identities on the brane imply that the projected Weyl tensor obeys
the constraint
\begin{equation}
\nabla^\mu{\cal E}_{\mu\nu}={6\kappa^2\over\lambda}\nabla^\mu
S_{\mu\nu}\,. \label{5}
\end{equation}
This shows how nonlocal bulk effects are sourced by local bulk
effects, which include spatial gradients and time derivatives:
evolution and inhomogeneity in the matter fields can generate
nonlocal gravitational effects in the bulk, which backreact on the
brane. The brane energy-momentum tensor and the consolidated
effective energy-momentum tensor are {\em both} conserved
separately. These conservation equations, as well as the brane
field equations and Bianchi identities, are given in covariant
form in~\cite{m}. We are interested here in the particular case of
a Bianchi~I brane geometry, the simplest anisotropic
generalization of an FRW brane geometry.

\section{Anisotropic brane}

A Bianchi~I brane has the induced metric
\begin{equation}\label{bianchi1}
ds^2 = -dt^2 + R_i^2(t) (dx^i)^2\,,
\end{equation}
and is covariantly characterized by
\begin{equation}\label{bianchi}
\D_\mu f=0\,,~A_\mu=0=\omega_\mu\,,~ {\cal
Q}_\mu=0\,,~R^*_{\mu\nu}=0\,,
\end{equation}
where $\D_\mu$ is the projected covariant spatial derivative, $f$
is any physically defined scalar, $A_\mu$ is the 4-acceleration,
$\omega_\mu$ is the vorticity, and $R^*_{\mu\nu}$ is the Ricci
tensor of the 3-surfaces orthogonal to $u^\mu$. (Note that in the
coordinates of Eq.~(\ref{bianchi1}), we have
$u_\mu=-\delta_\mu{}^0$, $h_{\mu 0}=0$, $\D_\mu
f=\delta_\mu{}^i\partial_i f$.)

The conservation equations~\cite{m} reduce to
\begin{eqnarray}
&&\dot{\rho}+\Theta(\rho+p)=0\,,\label{pc1}\\ && \dot{\cal
U}+{\textstyle{4\over3}}\Theta{\cal U}+\sigma^{\mu\nu}{\cal
P}_{\mu\nu}  =0 \,, \label{lc1'}\\&& \D^\nu{\cal P}_{\mu\nu}
=0\,,\label{lc2'}
\end{eqnarray}
where a dot denotes $u^\nu\nabla_\nu$, $\Theta$ is the volume
expansion rate, and $\sigma_{\mu\nu}$ is the shear. Introducing
the directional Hubble parameters $H_i = {\dot R_i}/R_i$ and the
mean expansion factor $a = (R_1R_2R_3)^{1/3}$, one gets $\Theta
\equiv 3H = 3{\dot a}/a \equiv \sum_{i} H_i$.

There is no evolution equation for ${\cal P}_{\mu\nu}$, reflecting
the fact that in general the equations do not close on the brane,
and one needs bulk equations to determine brane dynamics. There
are bulk degrees of freedom whose impact on the brane cannot be
predicted by brane observers.

The generalized Raychaudhuri equation on the brane~\cite{m}
becomes (with $\Lambda=0$)
\begin{eqnarray}
&&\dot{\Theta}+{\textstyle{1\over3}}\Theta^2
+\sigma^{\mu\nu}\sigma_{\mu\nu}
+{\textstyle{1\over2}}\kappa^2(\rho + 3p) \nonumber\\&&~~{}=
-{\textstyle{1\over2}} \kappa^2 (2\rho+3p){\rho\over\lambda} -{6
{\cal U}\over\kappa^2\lambda}\,, \label{prl}
\end{eqnarray}
where the general relativistic case is recovered when the
right-hand side is set to zero. The vanishing of $R^*_{\mu\nu}$
leads via the Gauss-Codazzi equations on the brane to
\begin{eqnarray}
\dot{\sigma}_{\mu\nu}+\Theta\sigma_{\mu\nu}&=&{6
\over\kappa^2\lambda}{\cal P}_{\mu\nu}\,,\label{g1}\\
-{\textstyle{2\over3}}\Theta^2 +\sigma^{\mu\nu}\sigma_{\mu\nu}
+2\kappa^2\rho &=& -\kappa^2{\rho^2\over\lambda} -{12 {\cal
U}\over\kappa^2\lambda}\,.\label{g2}
\end{eqnarray}

The presence of the nonlocal bulk tensor ${\cal P}_{\mu\nu}$ on
the right of Eq.~(\ref{g1}) means that we cannot simply integrate
to find the shear as in general relativity (see~\cite{e}).
However, we can circumvent this problem in a special case: when
the nonlocal energy density vanishes
or is negligible, i.e.,
\begin{equation}\label{u}
{\cal U}=0\,.
\end{equation}
This assumption is often made in the case of FRW branes, and in
that case, it leads to a conformally flat bulk
geometry~\cite{bdel}. When Eq.~(\ref{u}) holds, then the
conservation equation~(\ref{lc1'}) implies $\sigma^{\mu\nu}{\cal
P}_{\mu\nu}=0$.
This consistency condition implies a condition on the evolution of
${\cal P}_{\mu\nu}$, i.e., $\sigma^{\mu\nu}\dot{\cal
P}_{\mu\nu}=6{\cal P}^{\mu\nu} {\cal P}_{\mu\nu}/\kappa^2\lambda$,
as follows from Eq.~(\ref{g1}). Since there is no evolution
equation for ${\cal P}_{\mu\nu}$ on the brane~\cite{m}, this is
consistent on the brane. However, we would need to check that the
brane metric with ${\cal U}=0$ leads to a physical 5-dimensional
bulk metric. This would have to be done numerically (the bulk
metric for a Bianchi brane-world is not known), and is a topic for
further investigation.

Equation~(\ref{g1}) may be integrated after contracting it with
the shear, to give
\begin{equation}\label{s}
\sigma^{\mu\nu}\sigma_{\mu\nu} \equiv \sum_{i = 1}^3 (H_i - H)^2 =
{6\Sigma^2\over a^6}\,,~~\dot\Sigma=0\,.
\end{equation}
We now substitute into Eq.~(\ref{g2}) to obtain the generalized
Friedmann equation for the Bianchi~I brane (with $\Lambda=0={\cal
U}$):
\begin{equation}\label{f}
H^2={\kappa^2\over3} \rho\left(1 +{\rho\over2\lambda}\right) +
{\Sigma^2\over a^6}.
\end{equation}
When $\Sigma=0$, this recovers the equation for an FRW
brane~\cite{bdel}. When $\rho/\lambda\to 0$, we recover the
equation for a Bianchi~I model in general relativity~\cite{e}.

\begin{figure}[bth]
\includegraphics[width=3.3in]{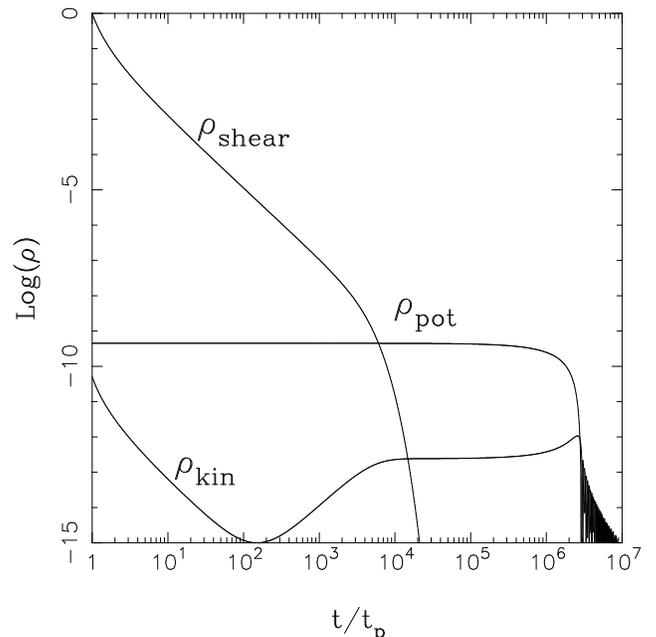}
\caption{This figure shows the evolution of the three components
contributing to the effective energy density, when
$V={1\over2}m\phi^2$. The universe evolves from $t=t_{\rm{p}}$ and
$a=1$, with initial conditions $\dot{\phi}= -10^{-5}$, $\phi = 3$,
$\rho/\lambda=5 $ and $\rho_{\rm shear} = 1$; we take $m =
10^{-5}$ ($M_{\rm p} = 1$ is assumed). The kinetic term initially
plunges under the influence of the shear and rises after the shear
stops dominating the dynamics of the universe. The oscillating
phase at the end of inflation shows up as the dense patch at the
right of the plot.}
\end{figure}

\section{Inflation on the anisotropic brane}
The evolution equation for a minimally coupled scalar field confined to
the brane is
\begin{equation}\label{kg}
{\ddot \phi} + 3H {\dot \phi} + V'(\phi) = 0.
\end{equation}
The energy density and pressure are respectively
\begin{equation}\label{kinpot}
\rho=\rho_{\rm kin}+\rho_{\rm pot} \,,~ p=\rho_{\rm kin}-\rho_{\rm
pot}\,.
\end{equation}
where $\rho_{\rm kin} = {1\over2}\dot\phi^2$, $\rho_{\rm
pot}=V(\phi)$. Setting $\Sigma = 0$ in Eq.~(\ref{f}), we see that
the extra-dimensional terms act to increase the Hubble rate, and
hence the damping experienced by the scalar field as it rolls down
its potential. Thus for a FRW brane, inflation at high energies
($\rho>\lambda$) proceeds at a higher rate than the corresponding
rate in general relativity. This introduces important changes to
the dynamics of the early universe~\cite{mwbh,cll,bdel}, and
accounts for an increase in the amplitude of scalar~\cite{mwbh}
and tensor~\cite{lmw} fluctuations at Hubble-crossing, and for a
change to the evolution of large-scale density perturbations
during inflation~\cite{gm}.

The condition for inflation is $\ddot{a}>0$, which is equivalent
to $\dot{\Theta}+{\textstyle{1\over3}}\Theta^2>0$. By
Eq.~(\ref{prl}), with ${\cal U}=0$, this gives
\begin{equation}\label{inf}
w<-{1\over3}\left({2\rho+\lambda\over \rho+\lambda} \right)-\left(
{2\over 1+\rho/\lambda}\right){\sigma_{\mu\nu}\sigma^{\mu\nu}\over
3\kappa^2 \rho} \,,
\end{equation}
where $w=p/\rho$. When the shear vanishes, this reduces to the
condition for FRW brane inflation given in~\cite{mwbh}; if
$\rho/\lambda\to\infty$, we have $w<-{2\over3}$, while the general
relativity condition $w<-{1\over3}$ is recovered as
$\rho/\lambda\to 0$. For the Bianchi~I brane, the condition
becomes
\begin{equation}\label{inf'}
w<-{1\over3}\left({2\rho+\lambda\over \rho+\lambda} \right)-
{4\Sigma^2\over \kappa^2 a^6\rho (1+\rho/\lambda)} \,.
\end{equation}

From Eqs.~(\ref{f}) and (\ref{inf'}), one might naively expect the
presence of shear to be detrimental for inflation, since: (i)
Eq.~(\ref{inf'}) implies that a more negative equation of state is
necessary to accelerate the universe in the presence of shear;
(ii) Eq.~(\ref{f}) suggests that a large initial value of the
anisotropy $\rho_{\rm shear}=3\Sigma^2/ \kappa^2a^6 \gg \rho_\phi$
might, by dominating the expansion dynamics of the early universe,
prevent inflation from occurring. A closer examination of the
situation however reveals both these arguments to be flawed. From
Eqs.~(\ref{f}) and (\ref{kg}), we see that the presence of
expansion anisotropy (shear) acts in a manner which is actually
{\em conducive} to inflation. For a Bianchi~I brane, the shear
anisotropy term in Eq.~(\ref{f}) reinforces the increase of the
Hubble rate. A larger value of initial shear damps the kinetic
energy of the scalar field, allowing the inflationary condition
Eq.~(\ref{inf'}) to be reached earlier. The important role played
by anisotropy is illustrated by considering the `worst case'
scenario for inflation when $\rho_{\rm kin}={\dot \phi}^2/2 \gg
\rho_{\rm pot}= V(\phi)$. If we assume that the universe was
initially very anisotropic, so that its expansion rate was
dominated by the shear term, we get
\[
a^3(t) \approx a_0^3 + 3\Sigma (t-t_0)\,.
\]
Consequently, at early times
\begin{equation}\label{shear1}
\rho_{\rm shear}\,, \rho_{\rm kin} \sim (t-t_0)^{-2}\,.
\end{equation}
This estimate is confirmed by numerical integration; see
Figs.~1--3. Remarkably, we find that anisotropy always disappears
within a fixed interval of time, no matter what its initial value
(compare~\cite{ms86}). The decrease in kinetic energy is
influenced by the value of the initial anisotropy; a larger value
of $\Sigma$ causes a more rapid decay of the kinetic term and
therefore results in an earlier onset of inflation.


\begin{figure}[tbh]
\includegraphics[width=3.3in]{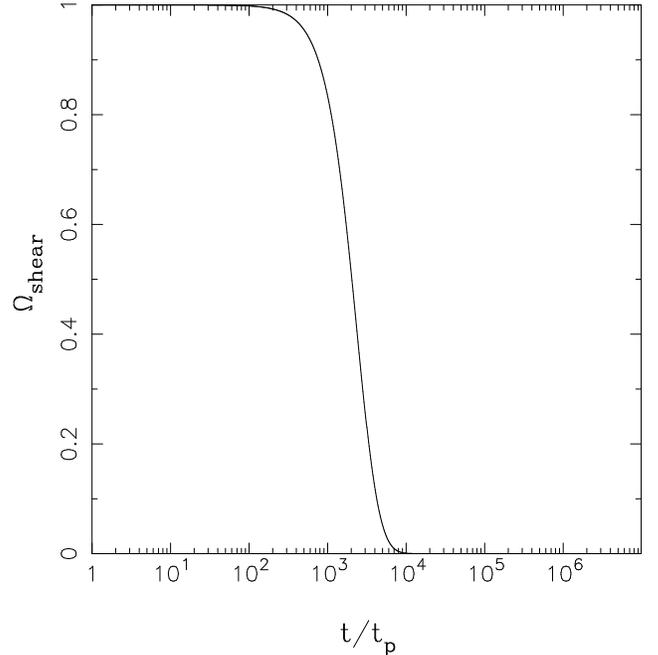}
\caption{The evolution of the dimensionless shear parameter
$\Omega_{\rm shear} = \sigma^2/6H^2$
is shown as a function of time for the model in Fig.~1.
The shear decreases monotonically as the universe expands
and isotropises.}
\end{figure}

\section{Conclusions}

Our results illustrated in Figs.~1 and 3 show the kinetic,
potential and `anisotropy' energy densities plotted as functions
of time. The associated dimensionless shear parameter $\Omega_{\rm shear}
= \sigma^2/6H^2$ is shown in Figs.~2 and 4 and the expansion factor in Fig. 5.

\begin{figure}[tbh]
\includegraphics[width=3.3in]{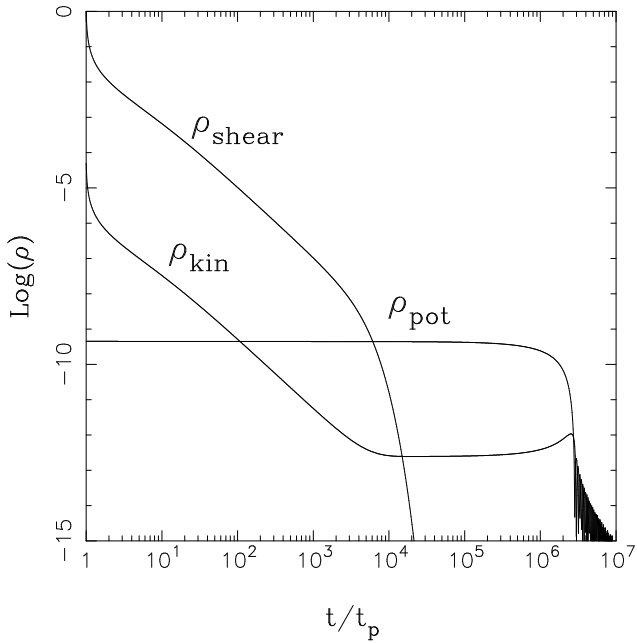}
\caption{Same as Fig.~1, but with a larger initial value of the
field velocity, $\dot{\phi} = -10^{-2}$ and $\rho/\lambda = 5 \times 10^5$. }
\end{figure}

We find that the influence of anisotropy on the kinetic energy is
particularly strong if $\Sigma^2/a^6\kappa^2 \gg {\dot\phi}^2$
initially. In this case the kinetic term drops to a very small
value, then rises after the anisotropy has disappeared, gradually
approaching its asymptotic `slow-roll' value
\begin{equation}\label{slow}
{\dot \phi} \simeq -V'/3H ~\Rightarrow ~ {\dot\phi}^2 \simeq
\frac{\lambda}{3\pi}\left(\frac{M_{\rm p}}{\phi}\right)^2\,.
\end{equation}
We assume here that the potential has the simple `chaotic' form $V
={1\over2} m^2\phi^2$, for which the standard inflationary
slow-roll condition is ${\dot\phi}^2 \simeq m^2 M_{\rm
p}^2/16\pi^2$. Comparing with Eq.~(\ref{slow}), we find that
dependence on $m^2$ has been replaced in the brane scenario by
$\lambda/\phi^2$. Thus the kinetic energy does not remain constant
but gradually increases as the field amplitude decreases during
slow-roll. We find that the decay of anisotropy is generically
accompanied by a corresponding decrease in the kinetic energy of
the scalar field. This effect leads to greater inflation; see
Fig.~5.

\begin{figure}[h]
\includegraphics[width=3.3in]{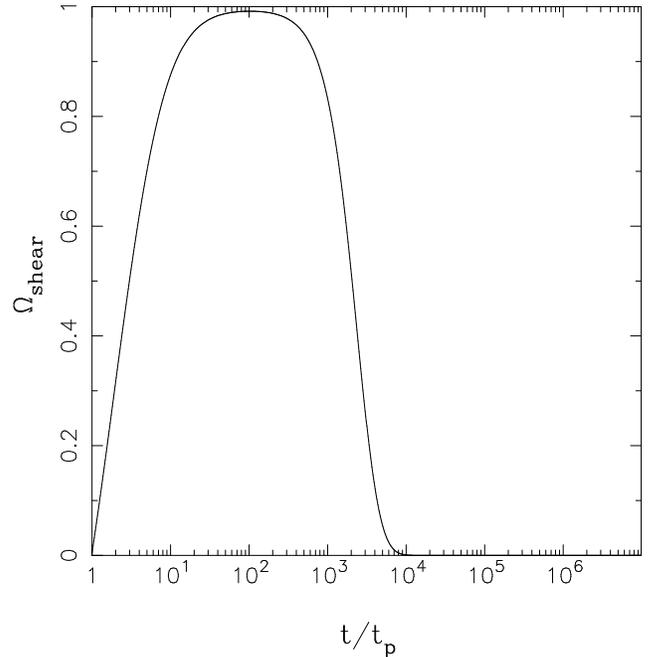}
\caption{The evolution of the dimensionless shear parameter
$\Omega_{\rm shear} = \sigma^2/6H^2$
is shown as a function of time for the model in Fig. ~3.
We find that the early and late-time expansion of the universe is isotropic,
but the shear term dominates
during an intermediate anisotropic stage.}
\end{figure}

Examining Eq.~(\ref{f}) closely
we find that the effective equation of state of matter when
$\rho \gg \lambda$ is
$w_{\rm eff} = 2w + 1$.
Consequently for matter with $w = p/\rho > 0$
the approach to the initial singularity is matter
dominated and not shear dominated, due to the predominance of the matter term
$\rho^2/2\lambda^2$ relative to the shear term
$\Sigma^2/a^6$. (Within the framework of
scalar field models this corresponds to ${\dot\phi}^2 > 2V(\phi)$.)
The fact that the density effectively grows faster than $1/a^6$ for
$w > 0$ is a uniquely brane effect. Within the framework of general relativity
such behaviour is clearly impossible since it would demand an ultra-stiff
equation of state $w > 1$.
The expansion of the early universe is therefore characterised by three
successive expansion epochs during which the Bianchi I model
experiences: (i) initial quasi-isotropic expansion $\sigma^2/H^2 \to 0,
t \to 0$; (ii) transient anisotropy dominated regime
$\sigma^2/H^2 \sim 1 $; (iii) anisotropy dissipation
$\sigma^2/H^2 \to 0, t \to \infty$. These three stages are illustrated
in Fig ~4. For smaller values of ${\dot\phi}^2$ (corresponding to
$w < 0$) stage (i) is absent and the shear decreases monotonically from
a large initial value; see Fig. ~2.

\begin{figure}[tbh]
\includegraphics[width=3.3in]{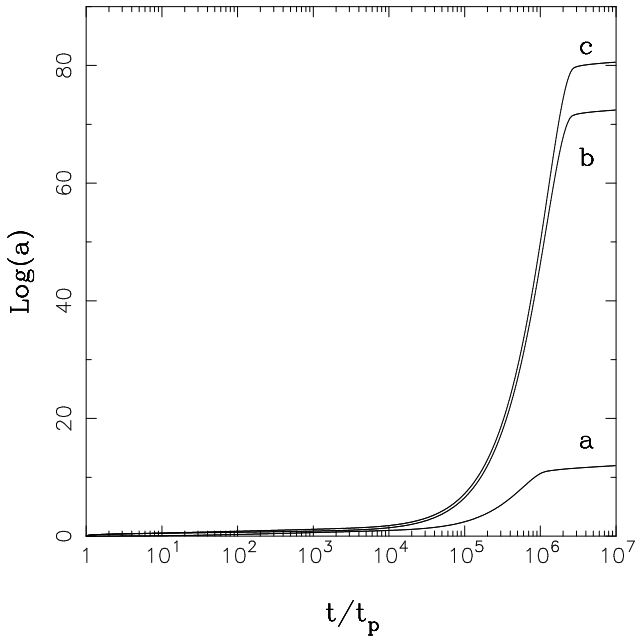}
\caption{The evolution of the scale factor with time is shown for
the model in Fig.~3. The curves shown in increasing amplitude from
bottom to top correspond to: (a) $\Sigma=0$, $1/2\lambda = 0$ (the
general relativity case); (b) $\Sigma=0$, $1/2\lambda = 10^{10}$
and (c) $\Sigma=1$, $1/2\lambda = 10^{10}$. We see that the effect
of both the extra-dimensional term and cosmic shear is to produce
more inflation.}
\end{figure}

The decay of shear and the accompanying isotropization of
the universe significantly increases the class of initial
conditions from which the present universe could have originated.

\[ \]
{\bf Acknowledgements:}\\

We would like to thank Alexei Starobinsky and
Alexey Toporenskij for helpful
comments which led to improvments in the manuscript.
Discussions with Naresh Dadhich
are also acknowledged.\\


\begin{references}
\bibitem[*]{byline} Electronic address: roy.maartens@port.ac.uk
\bibitem[\dagger]{byline} Electronic address: varun@iucaa.ernet.in
\bibitem[\ddagger]{byline} Electronic address: saini@iucaa.ernet.in

\bibitem{ch73}
C.B. Collins and S.W. Hawking, Astrophys. J. {\bf 180}, 317
(1973).

\bibitem{wald}
R.M. Wald. Phys. Rev. D {\bf 28}, 2118 (1983).

\bibitem{star83}
A.A. Starobinsky, JETP Lett. {\bf 37}, 66 (1983)

\bibitem{ms86}
I. Moss and V. Sahni, Phys. Lett. {\bf B178}, 159 (1986).

\bibitem{tw86}
M.S. Turner and L.M. Widrow, Phys. Rev. Lett. {\bf 57}, 2237 (1986).

\bibitem{js86}
L.G. Jensen and J. Stein-Schabes, Phys. Rev. D {\bf 34}, 831 (1986).

\bibitem{rellis86}
T. Rothman and G.F.R. Ellis, Phys. Lett. {\bf B180}, 19 (1986).

\bibitem{rs}
L. Randall and R. Sundrum, Phys. Rev. Lett. {\bf 83}, 4690 (1999).

\bibitem{mwbh}
R. Maartens, D. Wands, B.A. Bassett, and I.P.C. Heard, Phys. Rev.
D {\bf 62}, 041301 (2000).

\bibitem{cll}
J. Cline, C. Grojean, and G. Servant, Phys. Rev. Lett. {\bf 83},
4245 (1999);\\ C. Csaki, M. Graesser, C. Kolda, and J. Terning,
Phys. Lett. {\bf B462}, 34 (1999);\\ D. Ida, J. High Energy Phys.
{\bf 09}, 014 (2000);\\ R.N. Mohaptra, A. Perez-Lorenzana, and
C.A. Pires, Phys. Rev. D {\bf 62}, 105030 (2000) and
hep-ph/0003328;\\ L. Mendes and A.R. Liddle, Phys. Rev. D {\bf
62}, 103511 (2000);\\ E.J. Copeland, A.R. Liddle, and J.E. Lidsey,
astro-ph/0006421;\\ A. Mazumdar, hep-ph/0007269 and
hep-ph/0008087;\\ R.M. Hawkins and J.E. Lidsey, Phys. Rev. D {\bf
63}, 041301 (2001).

\bibitem{sms}
T. Shiromizu, K. Maeda, and M. Sasaki, Phys. Rev. D {\bf 62},
024012 (2000).

\bibitem{m}
R. Maartens, Phys. Rev. D {\bf 62}, 084023 (2000).

\bibitem{e} G.F.R. Ellis, in {\em Cargese Lectures}, ed. E. Shatzman
(Gordon \& Breach, 1973).

\bibitem{bdel}
P. Binetruy, C. Deffayet, U. Ellwanger, and D. Langlois, Phys.
Lett. {\bf B477}, 285 (2000).

\bibitem{lmw}
D. Langlois, R. Maartens, and D. Wands, Phys. Lett. {\bf B489},
259 (2000).

\bibitem{gm}
C. Gordon and R. Maartens, Phys. Rev. D {\bf 63}, 044022 (2001).

\end{references}
\end{document}